\documentclass[a4paper,10pt]{scrartcl}
\usepackage{amsmath}
\usepackage{amsfonts}
\usepackage{amssymb}
\usepackage{graphicx}
\usepackage[english]{babel}
\usepackage{wasysym}
\usepackage{units}
\newcommand{\mt}[1]{\mathcal{#1}}
\newcommand{\p}{\partial}
\newcommand{\q}{\quad}
\newcommand{\qq}{\qquad}
\newcommand{\ph}{\phantom1}
\newcommand{\reff}[1]{(\ref{#1})}
\newcommand{\vs}[1]{\vspace{#1mm}}
\newcommand{\vsO}{\vspace{.1cm}\hfill\\}
\newcommand{\vsT}{\vspace{.2cm}\hfill\\}

\title{\Large FERMION DYNAMICS BY INTERNAL\\AND SPACE-TIME SYMMETRIES}

\author{{\large N. Carlevaro}$^{\;a,\,b}$, {\large O.M. Lecian}$^{\;a,\,c}$ 
{\large and G. Montani}$^{\;a,\,c,\,d,\,e}$\vsT
\emph{\footnotesize $^a$ICRA -- International Center for Relativistic Astrophysics,}\vs{-2.5}\\
\emph{\footnotesize c/o Dep. of Physics - ``Sapienza'' Universit\`a di Roma}\\
\emph{\footnotesize $^b$Department of Physics, Polo Scientifico -- Universit\`a degli Studi di Firenze,}\vs{-2.5}\\
\emph{\footnotesize INFN -- Section of Florence, Via G. Sansone, 1 (50019), Sesto Fiorentino (FI), Italy}\\
\emph{\footnotesize $^c$ Department of Physics - ``Sapienza'' Universit\`a di Roma, Piazza A. Moro, 5 (00185), Rome, Italy}\\
\emph{\footnotesize $^d$ENEA -- C.R. Frascati (Department F.P.N.), Via Enrico Fermi, 45 (00044), Frascati (Rome), Italy}\\
\emph{\footnotesize $^{e}$ ICRANet -- C. C. Pescara, Piazzale della Repubblica, 10 (65100), Pescara, Italy}\\
{\footnotesize\ttfamily nakia.carlevaro@icra.it\quad lecian@icra.it\quad montani@icra.it}
}
\date{}
\begin{document}
\maketitle

%
\hrule
\begin{abstract} \textbf{Abstract:} This manuscript is devoted to introduce a gauge theory of the Lorentz Group based on the ambiguity emerging in dealing with isometric diffeo-morphism-induced Lorentz transformations. The behaviors under local transformations of fermion fields and spin connections (assumed to be ordinary world vectors) are analyzed in flat space-time and the role of the torsion field, within the generalization to curved space-time, is briefly discussed. The fermion dynamics is then analyzed including the new gauge fields and assuming time-gauge. Stationary solutions of the problem are also analyzed in the non-relativistic limit, to study the spinor structure of an hydrogen-like atom.

\vsO \emph{PACS Nos.}: 02.40.-k; 11.30.Cp
\end{abstract}
\hrule

\vspace{1cm}
\section{General Remarks}
Many attempts to construct a gauge model of gravitation exist. In particular, the works by Utiyama \cite{utiyama56} and by Kibble \cite{kibble61} were the starting points for various gauge approaches to gravitation (as discussed in Section 2). As a result, Poincar\'e gauge theory (PGT), see \cite{hehl-von74,hehl-von76,blago1,blago2,deser}, is a generalization of the Einstein scheme of gravity, in which not only the energy-momentum tensor, but also the spin of matter plays a dynamical role when coupled to spin connections, in a non-Riemannian space-time.

To include spinor fields consistently, it is necessary to extend the framework of General Relativity (GR), as already realized by Hehl \emph{et al.} \cite{hehl-von76}: this necessity is strictly connected with the non existence in GR of an independent concept of spin momentum for physical fields, as the Lorentz Group (LG) has not an independent status of gauge group in GR. In fact, we will demonstrate that an isometric diffeomorphism can induce a local Lorentz rotation, thus standard spin connections $\omega_\mu^{\ph ab}$ have no longer a gauge role in this framework (being only a function of tetrads and behaving like vectors under the diffeomorphism-induced Lorentz rotation). New gauge connections $A_\mu^{\ph ab}$ have to be introduced into the dynamics to appropriately recover the Lorentz invariance of the scheme, when spinor fields are taken into account (as discussed in Section 3).

This paradigm is well established in flat space-time, where isometric diffeomorphism are allowed and spin connections can be set to zero. Within this framework, fermion dynamics is analyzed including the effects of the new gauge fields (as discussed in Section 4). A Modified Dirac Equation is the starting point to study the non-relativistic limit and the resulting Generalized Pauli Equation in presence of a Coulomb central potential. For an hydrogen-like atom, energy-level splits are predicted but no new spectral lines are allowed (as discussed in Section 5).

The analysis developed in flat space-time can be extended to a curved one. In the First-Order Approach, the geometrical identification of the LG gauge fields with a suitable bein projection of the contortion field is allowed when a non-standard interaction term between generalized connections and these gauge fields is postulated, if fermion matter is absent. On the other hand, when spinors are present, both spin connections and the spin current contribute to the torsion term (as discussed in Section 6).\vspace{0.3cm}

{\textbf{\emph{Notation:}} Greek indices (\emph{e.g.}, $\mu=0,1,2,3$) change as tensor ones under general coordinate transformations (\emph{i.e.}, world transformations); Latin indices (\emph{e.g.}, $a=0,1,2,3$) are the tetradic indices and refer to Lorentz transformations; \emph{Only} indices $i,\;j,\;k$ are 3-dimensional indices and run from 1 to 3.}

\section{Internal and Space-Time Symmetries} This Section is aimed at analyzing the internal symmetries of the space-time. We focus on the description of GR as a gauge model, underling the ambiguity that arises from this approach.

We can introduce the usual orthonormal basis $e^{\ph a}_{\mu}$ (tetrads) for the local Minkowskian tangent space-time of a 4-dimensional manifold. Tetrads are locally defined in curved space-time and their transformations can be read as generic reference-system changes. This way, such a standard formalism allows one to recover Lorentz symmetry because tetrad changes are defined as local Lorentz transformations linking the different inertial references they describe.
\newcommand{\ls}{\scriptscriptstyle{(\omega)}}
\newcommand{\sss}{\scriptscriptstyle{(\omega)}}
The relations between tetrads and the metric $g_{\mu\nu}$ are
\begin{equation}\label{relation}
g_{\mu\nu}=\eta_{ab}\,e_{\mu}^{\ph a}\,e_{\nu}^{\ph b},\qquad\q
e_{\mu}^{\ph a}\,e^{\mu}_{\ph b}=\delta^{a}_{b},\qquad\q
e_{\mu}^{\ph a}\,e^{\nu}_{\ph a}=\delta^{\nu}_{\mu}\;,
\end{equation}
where $\eta_{ab}$ is the local Minkowski metric. Projecting tensor fields from the 4-dimensional manifold to the Minkowskian space-time allows us to emphasize the local Lorentz invariance of the scheme in presence of spinor fields. In fact, fermions transform like a particular representation $S$ of the LG, \emph{i.e.}, $\psi\to S\psi$, where
\begin{align}\label{LG}
S=I-\tfrac{i}{4}\;\epsilon^{{a}{b}}\,\Sigma_{{a}{b}}\;,\qq\qq
\Sigma_{ab}=\tfrac{i}{2}\,[\gamma_{a},\gamma_{b}]\;,\qq\qq
[\Sigma_{cd},\Sigma_{ef}]=i\mathcal{F}^{ab}_{cdef}\,\Sigma_{ab}\;,
\end{align}
here the $\Sigma_{ab}$'s and the $\mathcal{F}^{ab}_{cdef}$'s are the generators and the structure constants of the LG, respectively and $\epsilon^a_b(x)$ is the infinitesimal Lorentz rotational parameter. To assure the Lorentz covariance of the spin derivative $\p_\mu\,\psi$, connections $\omega_{\mu}^{\ph ab}$ must be introduced to define a covariant derivative as 
\begin{equation}\label{spin_connections}
D^{\sss}_\mu=\p_\mu+\Gamma^{\sss}_\mu\;,\qquad
\Gamma^{\sss}_\mu=\tfrac{1}{2}\;\omega_{\mu}^{\ph ab}\, \Sigma_{ab}\;,\qquad
\omega_{\mu}^{\ph ab}=e^{a\nu}\nabla_\mu e^{\ph b}_{\nu}=
e^{\ph c}_{\mu}\,\gamma^{ba}_{\ph\ph c}\;,
\end{equation}
where the $\omega_{\mu}^{\ph ab}$'s denote the so-called \emph{spin connections} and $\gamma_{abc}=e^{\mu}_{\ph c} e^{\nu}_{\ph b} \nabla_\mu e_{\nu a}$ are the Ricci Rotation Coefficients ($\nabla_\mu$ is the usual coordinate covariant derivative). It is worth underlining that the introduction of the tetrad formalism enables us to include spinor fields in the dynamics \cite{lusanna}. In this sense, spin connections are introduced to restore the correct Dirac algebra in curved space-time \emph{i.e.}, $D^{\sss}_\mu\;\gamma^{\nu}=0$ \cite{hammond}. By other words, the correct treatment of spinors in curved space-time leads to the introduction of those connections, which guarantee an appropriate gauge model for the LG.

This picture suggests, in appearance, the description of gravity as a gauge model \cite{cho1,cho2}. Spin connections are a bein projection of Ricci Rotation Coefficients and this formalism leads to the usual definition of the curvature tensor:
\begin{equation}\label{Riemann}
R^{\ph\ph ab}_{\mu\nu}=\p_\nu\omega_{\mu}^{\ph ab}-
\p_\mu\omega_{\nu}^{\ph ab}+\mt{F}^{ab}_{cdef}\omega_{\mu}^{\ph cd}\omega_{\nu}^{\ph ef}\;,
\end{equation}
which corresponds to the I Cartan Structure Equation. The Hilbert-Einstein Action rewrites now as
\begin{equation}\label{action for o}
S_G(e,\omega)=-\tfrac{1}{4}{\textstyle \int}
\mathrm{det}(e)\,d^{4}x\;\;e^{\ph\mu}_{a}e^{\ph\nu}_{b} R^{\ph\ph ab}_{\mu\nu}\;.
\end{equation}
Variation wrt connections leads to the II Cartan Structure Equation,
\begin{equation} \label{Cartan eq}
\p_\mu e^{\ph a}_{\nu} -\p_\nu e^{\ph a}_{\mu}-\omega_{\mu}^{\ph ab}e_{\nu b}
+\omega_{\nu}^{\ph ab}e_{\mu b}=0\;,
\end{equation}
which links the tetrad fields to the spin connections. Since $\omega_{\mu}^{\ph ab}=e^{\ph c}_{\mu}\,\gamma^{ba}_{\ph\ph c}$, we underline that such connections behave like ordinary vectors under general coordinate transformations (\emph{i.e.}, world transformations), and variation wrt tetrads gives rise the Einstein Equations, once the solution of eq. (\ref{Cartan eq}) is addressed.

In the standard approach, spin connections transform like \emph{Lorentz gauge vectors} under infinitesimal local Lorentz transformations (described by the Lorentz matrix $\Lambda_{a}^{b}=\delta_{a}^{b}+\epsilon_{a}^{b}$):
\begin{equation}\label{gaugetr}
\omega_{\mu}^{\ph ab}\stackrel{L}{\to} \omega_{\mu}^{\ph ab}-\p_\mu\epsilon^{ab}+
\tfrac{1}{4}\mt{F}^{ab}_{cdef}\epsilon^{cd}\omega_{\nu}^{\ph ef}
\end{equation}
and the Riemann tensor is preserved by such a change; therefore, in flat space-time, we deal with non-zero gauge connections, but a vanishing curvature. In both flat and curved space-time, the connections $\omega_{\mu}^{\ph ab}$ exhibit the right behavior to play the role of Lorentz gauge fields and GR exhibits the features of a gauge theory. On the other hand, the presence of tetrad fields (introduced by the Principle of General Covariance) is an ambiguous element for the gauge paradigm. In fact, spin connections can be uniquely determined as functions of tetrads in terms of the Ricci Rotation Coefficients $\gamma_{abc}$. This relation generates an ambiguity in the interpretation of the $\omega_{\mu}^{\ph ab}$'s as the only fundamental fields of the gauge scheme since the theory were based on two dependent degrees of freedom.

It is just the introduction of fermions that requires to treat local Lorentz transformations as the real independent gauge of GR. In fact, when spinor fields are taken into account, their transformations under the local Lorentz symmetry imply that the Dirac Equation is endowed with non-zero spin connections, even in flat space-time. Because of the behavior of spinor fields, it becomes crucial to investigate whether diffeomorphisms can be reinterpreted to some extent as local Lorentz transformations.

\section{A Novel Approach for a Gauge Theory of the LG} Here we want to fix some guidelines that are at the ground of this approach to Lorentz gauge theory. We first note that spin connections $\omega_{\mu}^{\ph ab}$ are gauge potentials and not physical fields, in the sense that they are subjected to such gauge transformations \reff{gaugetr} that do not alter the curvature tensor. 

The \emph{key point} is that, if we are able to show (as we will demonstrate in the following) that diffeomorphisms can induce local Lorentz transformations, we can conclude that the $\omega_{\mu}^{\ph ab}$'s can no longer be regarded as gauge potentials for the LG, because of the ambiguity of their transformation properties: do they behave like gauge fields or ordinary (coordinate) vectors? In this sense, new gauge fields must be added to retore the Lorentz invariance of the theory.

By other words, an inconsistence arises as far as the spinor behavior is analyzed under diffeomorphism-induced Lorentz rotation. Fermions are coordinate scalars and transform as the usual laws under Lorentz rotations, \emph{i.e.}, a spinor representation of the LG. They live in the tangent bundle without experiencing coordinate changes and, if the two transformations overlap, an ambiguity on the nature of spinors comes out. 

In flat space-time, in the case $e^{\ph a}_{\mu}=\delta^{\ph a}_{\mu}$, spin connections vanish and they must remain identically zero under local Lorentz transformations. In fact, the coordinate transformations can now be reinterpreted as Lorentz rotations and the request for external Lorentz gauge field appears well grounded. This picture remains still valid in curved space-time, as long as we accept that \emph{spin connections transform like vectors when diffeomorphism-induced rotations are implemented}. In this scheme, new gauge fields, transforming according to their Lorentz indices, are the only fields able to restore Lorentz invariance when local rotations are induced by coordinate changes. In fact, the nature of gauge potentials is naturally lost by the gravitational connections, which behave like tensors only. It is worth noting that, if the $\omega_{\mu}^{\ph ab}$'s are assumed to behave like gauge vectors under the induced Lorentz transformation, the standard approach can be recovered and the ambiguity of the tetrad dependence arises. It can also be interesting to compare our approach with that of \cite{ortin, kosmann}, where the formalism of the \emph{Lie Derivative} is extended to spinor fields.
\newcommand{\primo}{{\scriptscriptstyle\prime}}

We are now going to demonstrate that such correspondence between coordinate transformations and local rotations takes place only if we deal with \emph{isometric diffeomorphisms}. This request is naturally expected if we want to reproduce a Lorentz symmetry: an isometric diffeomorphism induces orthonormally-transformed basis and, to this extent, an isometry generates a local Lorentz transformation of the basis. An infinitesimal isometric diffeomorphism is described by
\begin{equation}\label{newdiff}
x^{\mu}\to x^{\primo\mu}=x^{\mu}+\xi^{\mu}(x)\;,\qq
\qquad\;\;\nabla_{\mu}\xi_{\nu}+\nabla_{\nu}\xi_{\mu}=0\;,
\end{equation}
and it induces the following transformation of the basis vectors
\begin{equation}\label{taylor}
e_{\mu}^{\ph a}(x)\stackrel{D}{\to}e_{\mu}^{\primo\ph a}(x^{\primo})=
e_{\nu}^{\ph a}(x)\;\nicefrac{\p x^{\nu}}{\p x^{\primo\mu}}=
e^{\ph a}_{\mu}(x)-e^{\ph a}_{\nu}(x)\;\nicefrac{\p \xi^{\nu}}{\p x^{\primo\mu}}\;.
\end{equation}
If we deal with an infinitesimal local Lorentz transformation $\Lambda_{a}^{b}(x)=\delta^b_a +\epsilon^{b}_{a}(x)$, we get, up to the leading order, the tetrad change (evaluated in $x^{\primo}$ of eq. \reff{newdiff}):
\begin{equation}
e_{\mu}^{\ph a}(x)\stackrel{L}{\to}e_{\mu}^{{\scriptscriptstyle \prime}\ph a}(x^{{\scriptscriptstyle \prime}})=\Lambda_{a}^{b}(x^{\scriptscriptstyle \prime})e_{\mu}^{\ph a}(x^{{\scriptscriptstyle \prime}})=
e_{\mu}^{\ph a}(x^{{\scriptscriptstyle \prime}})+e_{\mu}^{\ph b}(x)\epsilon^{a}_{b}(x)\;.
\end{equation}
In this sense, we can infer that the two transformation laws overlap if we assume the condition: 
\begin{equation}
\epsilon_{ab}=\nabla_{[a}\xi_{b]}-\gamma_{abc}\xi^c\;,
\end{equation}
where, to pick up local Lorentz transformations from the set of generic diffeomorphisms, the isometry condition $\nabla_{(\mu}\xi_{\nu)}=0$ has to be taken into account in order to get the antisymmetry condition $\epsilon_{ab}=-\epsilon_{ba}$ for the infinitesimal parameter $\epsilon_{ab}$.

\section{Spinors and Gauge Theory of the LG in Flat Space-Time} 
Let us now analyze the formulation of a gauge model for the LG in a flat Minkowski space-time when diffeomorphism-induced Lorentz transformation are allowed. The choice of flat space is due to the fact that the Riemann curvature tensor vanishes and, consequently, the usual spin connections $\omega_\mu^{\ph ab}$ can be set to zero choosing the gauge $e_\mu^{\ph a}=\delta_\mu^{\ph a}$ $^{(}$\footnote{In general, the $\omega_\mu^{\ph ab}$'s are allowed to be non-vanishing quantities.}$^{)}$. This allows one to introduce new Lorentz connections $A_\mu^{\ph ab}$ as the gauge fields as far as the correspondence between an infinitesimal diffeomorphism and a local Local rotation is recovered, as shown in the previous section.

In a 4-dimensional flat manifold, the metric tensor reads $g_{\mu\nu}=\eta_{ab}e^{\ph a}_{\mu}e^{\ph b}_{\nu}$ and an infinitesimal diffeomorphism and a local Lorentz transformation write as
\begin{equation}
x^a\stackrel{D}{\to}x^a +\xi^a (x^c)\;,\qquad\qq
x^a\stackrel{L}{\to}x^a +\epsilon^a_{b}(x^c)\;x^b\;,
\end{equation}
respectively. If vectors are taken into account, no inconsistencies arise if the two transformation coincide. In fact, if we set $\epsilon_a^b\equiv \p^b\xi_a(x^c)$, the two transformation laws
\begin{align}
^{D}V^{\primo}_a (x^{\primo})= V_a(x)+\p_a \xi^b(x)\;V_b(x)\;,\qq\q
^{L}V^{\primo}_a (x^{\primo})= V_a(x)+\epsilon^b_a(x)\;V_b(x)\;,
\end{align}
overlap and the LG loses its status of independent gauge group \cite{hehl-von76}. Here, the isometry condition $\p_b\xi_a+\p_a\xi_b=0$ has to be imposed to restore the proper number of degrees of freedom of a Lorentz transformation, \emph{10}, out of that of a generic diffeomorphism, \emph{16}. 

On the other hand, spin$-\nicefrac{1}{2}$ fields are described by the usual Lagrangian density
\begin{equation}
\mathcal{L}_F=\tfrac{i}{2}\;\bar{\psi}\gamma^ae^{\mu}_{\ph a}\p_{\mu}\psi-
\tfrac{i}{2}\;e^{\mu}_{\ph a}\p_{\mu}\bar{\psi}\gamma^a\psi
\end{equation}
and, if accelerated coordinates are taken into account, they have to recognize the isometric components of the diffeomorphism as a local Lorentz transformation, differently from vectors: a spinor can noway be a Lorentz scalar. In this scheme, new Lorentz connections $A_\mu^{\ph ab}$ have to be introduced for matter fields since, for assumption, the standard connections $\omega_{\mu}^{\ph ab}$ do not follow Lorentz gauge transformations and they are not able to restore Lorentz invariance. Let us implement an infinitesimal local Lorentz transformation considering the spin$-\nicefrac{1}{2}$ representation $S=S(\Lambda(x))$ of eqs. \reff{LG}, \emph{i.e.},
\begin{align}
S=I-\tfrac{i}{4}\;\epsilon^{{a}{b}}\,\Sigma_{{a}{b}}\;,\qq\qq
\Sigma_{ab}=\tfrac{i}{2}\,[\gamma_{a},\gamma_{b}]\;,\qq\qq
[\Sigma_{cd},\Sigma_{ef}]=i\mathcal{F}^{ab}_{cdef}\,\Sigma_{ab}\;.
\end{align}\newcommand{\as}{\scriptscriptstyle{(A)}}Such a transformation acts on the spinor in the standard way $\psi(x)\to S\;\psi(x)$ and $\gamma$ matrices are assumed to transform like Lorentz vectors, \emph{i.e.},
$S\,\gamma^{{a}}\,S^{-1}=(\Lambda^{-1})^{{a}}_{{b}}\,\gamma^{{b}}$. In this approach, gauge invariance is restored by a new covariant derivative, \emph{i.e.}, $\p_\mu\to D^{\as}_\mu$,
\begin{equation}
D^{\as}_\mu\psi=(\p_\mu-\tfrac{i}{4}\,A_\mu)\,\psi=
(\p_\mu-\tfrac{i}{4}\,A_\mu^{\ph ab}\,\Sigma_{ab})\,\psi\;,
\end{equation}
which behaves correctly like $\gamma^{\mu}D^{\as}_{\mu}\psi\to S(\Lambda)\gamma^{\mu}D^{\as}_{\mu}\psi$, since the field $A_\mu=A^{\ph ab}_\mu\Sigma_{ab}$ transforms under the following law:
$A_\mu\rightarrow S\,A_\mu\,S^{-1}-4i\,S\,\p_\mu\,S^{-1}$. Connections $A_\mu^{\ph ab}\,\,(\neq\omega_\mu^{\ph ab})$ transform like
\begin{equation}
A_\mu^{\ph ab}\rightarrow A_\mu^{\ph ab}-\p_\mu\epsilon^{ab}+4\mathcal{F}^{\,ab}_{cdef}\,\, \epsilon^{ef}\,A^{\ph cd}_\mu\;,
\end{equation}
\emph{i.e.}, as natural Yang-Mill fields associated to the LG, living in the tangent bundle $^{(}$\footnote{The tangent bundle coordinates differ for the presence of an infinitesimal displacement from those of the Minkowskian space \cite{lecian}.}$^{)}$. A Lagrangian associated to the gauge connections can be constructed by the introduction of the gauge field strength
\begin{equation}
F_{\mu\nu}^{\ph\ph ab}=\p_\mu A_\nu^{\ph ab}-\p_\nu A_\mu^{\ph ab}
+\tfrac{1}{4}\mt{F}^{ab}_{cdef}A_\mu^{\ph cd}A_\nu^{\ph ef}\;,
\end{equation}
which is not invariant under gauge transformations, as usual in Yang-Mills gauge theories, but the gauge invariant Lagrangian for the model
\begin{equation}\label{action-for-A}
\mt{L}_A=-\tfrac{1}{4}\;F_{\mu\nu}^{\ph\ph ab}F^{\mu\nu}_{\ph\ph ab}\;,
\end{equation}
can be introduced. In flat space, the only real dynamical fields are the Lorentz gauge fields. Variation of the total action, derived by the total lagrangian density $\mt{L}_{tot}=\mt{L}_F(D^{\as}_\mu\psi)+\mt{L}_A$, wrt the new gauge connections leads to the dynamical equations
\begin{equation}
\nabla_{\mu}F^{\mu\nu}_{\ph\ph ab}= J^{\nu}_{\ph ab}\;,\qquad\qq
J^{\nu}_{\ph ab}=-\tfrac{1}{4}\,\epsilon^{cd}_{ab}e_{\ph c}^{\nu}\,j^{\,(ax)}_d\;,
\end{equation}
where $j^{\,(ax)}_d=\bar{\psi}\,\gamma_5\gamma_d\,\psi$ is the spin axial current. Such field equations correspond to the Yang-Mills Equations for the non-Abelian gauge fields of the LG in flat space-time. The source of these gauge fields is the conserved spin density of the fermion matter whose dynamics will be analyzed in the next section.

\newcommand{\mj}{m_{\textrm{\tiny$j$}}}
\newcommand{\ml}{m_{\textrm{\tiny$\ell$}}}
\newcommand{\ms}{m_{\textrm{\tiny$s$}}}

\newcommand{\basel}{\mid\!n;\,\ell\,\ml\,s\,\ms\rangle}
\newcommand{\BRAbasel}{\langle n;\,\ell\,\ml\,s\,\ms\!\mid}

\newcommand{\baselpr}{\mid\!n';\,\ell'\,\ml'\,s\,\ms'\rangle}
\newcommand{\BRAbaselpr}{\langle n';\,\ell'\,\ml'\,s\,\ms'\!\mid}

\newcommand{\basej}{\mid\!n;\,\ell\,s\,j\,\mj\rangle}
\newcommand{\BRAbasej}{\langle n;\,\ell\,s\,j\,\mj\!\mid}

\newcommand{\basejpr}{\mid\!n';\,\ell'\,s\,j'\,\mj'\rangle}
\newcommand{\BRAbasejpr}{\langle n';\,\ell'\,s\,j'\,\mj'\!\mid}

\newcommand{\balpha}{\mid\!\alpha;\,j\,\mj\rangle}
\newcommand{\BRAbalpha}{\langle\alpha;\,j\,\mj\!\mid}

\newcommand{\balphapr}{\mid\!\alpha';\,j'\,\mj'\rangle}
\newcommand{\BRAbalphapr}{\langle\alpha';\,j'\,\mj'\!\mid}

\section{Generalized Pauli Equation}
The aim of this Section is investigating the effects that the new gauge fields can generate in a flat space-time. In particular, we treat the interaction between the new connections $A_\mu^{\ph ab}$ and the 4-spinor $\psi$ (of mass $m$) to generalize the well-known Pauli Equation, which corresponds to the motion equation of an electron in presence of an electro-magnetic field \cite{shapiro02}. 

The implementation of the diffeomorphism-induced local Lorentz symmetry ($\partial_{\mu}\to D^{\as}_{\mu}$) in flat space, leads to the fermion Lagrangian density
\begin{equation}\label{lagrangian-tot}
\mathcal{L}_F=\tfrac{i}{2}\;\bar{\psi}\gamma^a e^{\mu}_{\ph a}\p_{\mu}\psi-
\tfrac{i}{2}\;e^{\mu}_{\ph a}\p_{\mu}\bar{\psi}\gamma^a\psi\,-\,m\,\bar{\psi}\psi\;+\;
\tfrac{1}{8}\,e^{\mu}_{\ph c}\,\bar{\psi}\,\{\gamma^{c},
\Sigma_{ab}\}\,A^{\ph ab}_{\mu}\,\psi\;,
\end{equation}
where $\{\gamma^c,\Sigma_{ab}\}=2\,\epsilon^{c}_{abd}\,\gamma_5\,\gamma^d$. To study the interaction term, let us now start from the explicit expression 
\begin{equation}
\mathcal{L}_{int}=\tfrac{1}{4}\;\bar{\psi}\;
\epsilon^{c}_{abd}\,\gamma_5\,\gamma^d\,A^{ab}_{c}\;\psi\;.
\end{equation}
Let us now consider the role of gauge fields by analyzing its components $A^{0i}_{0}\,,\;A^{ij}_{0}\,,\;A^{0i}_{k}\,,
\;A^{ij}_{k}$, imposing the \emph{time-gauge} condition $A^{ij}_{0}\;=0$ associated to this picture and neglecting the term $A^{0i}_{0}\,$ since it summ over the completely anti-symmetric symbol $\epsilon^{0}_{0i d}\equiv0$. The interaction Lagrangian density rewrites now
\begin{equation}\label{lagrangian-split}
\mathcal{L}_{int}=\;\psi^{\dagger}\,C_0\,\gamma^{0}\gamma_5\gamma^0\,\psi\;+
\;\psi^{\dagger}\,C_i\,\gamma^{0}\gamma_5\gamma^i\,\psi\;,
\end{equation}
with the following identifications
\begin{equation}
C_0=\tfrac{1}{4}\;\epsilon^{k}_{ij0}A^{ij}_{k}\;,\qquad\quad
C_i=\tfrac{1}{4}\;\epsilon^{k}_{0ji}A^{0j}_{k}\;.
\end{equation}
Here the component $C_0$ is related to rotations, while $C_i$ to the boosts. Varying now the total action built up from the fermion Lagrangian density wrt $\psi^{\dagger}$, we get the \emph{Modified Dirac Equation} 
\begin{equation}\label{modified-dirac}
(i\,\gamma^0\gamma^0\p_0\;+\;
C_i\,\gamma^0\gamma_5\gamma^i\;+
\;i\,\gamma^0\gamma^i\p_i\;+\;
C_0\,\gamma^0\gamma_5\gamma^0)\,\psi\;=
\;m\,\gamma^0\,\psi\;,
\end{equation}
which governs the 4-spinor $\psi$ interacting with the new Lorentz gauge fields described here by the fields $C_0$ and $C_i$.

Let us now look for stationary solutions of the Dirac Equation expanded as
\begin{equation}\nonumber
\psi(\textbf{r},t)\to\psi(\textbf{r})\;e^{-i\mathcal{E}t}\;,\qquad\quad
\psi=\left(\begin{array}{l}\!\chi\!\\\!\phi\!\end{array}\right)\;,\qquad\quad
\psi^\dagger=(\,\chi^\dagger\;,\;\phi^\dagger\,)\;,
\end{equation}
where $\mathcal{E}$ denotes the spinor total energy and the 4-component spinor $\psi(\textbf{r})$ is expressed in terms of the two 2-spinors $\chi(\textbf{r})$ and $\phi(\textbf{r})$. Using now the standard representation of the Dirac matrices, the modified Dirac Equation \reff{modified-dirac} splits into two coupled equations (here we write explicitly the \emph{3}-momentum $p^{\,i}$):
\begin{subequations}\label{stationary-eq}
\begin{align}
(\mathcal{E}-\sigma_i\,C^i)\,\chi\;
-\;(\sigma_i\,p^{\,i}+C_0)\,\phi\;&=\;m\,\chi\;,\label{stationary-eq1}\\
(\mathcal{E}-\sigma_i\,C^i)\,\phi\;
-\;(\sigma_i\,p^{\,i}+C_0)\,\chi\;&=-\;m\,\phi\;.\label{stationary-eq2}
\end{align}
\end{subequations}

Let us now investigate the non-relativistic limit by splitting the spinor energy in the form
$\mathcal{E}=E+m$. Substituting this expression in the system \reff{stationary-eq}, we note that both the $|E|$ and $|\,\sigma_i\,C^i|$ terms are small in comparison wrt the mass term $m$ in the low-energy limit. Then, eq. \reff{stationary-eq2} can be solved approximately as
\begin{equation}\label{small-components}
\phi\;=\;\tfrac{1}{2m}\;(\sigma_i\,p^{\,i}+C_0)\,\chi\;.
\end{equation}
It is immediate to see that $\phi$ is smaller than $\chi$ by a factor of order $\nicefrac{p}{m}$ (\emph{i.e.}, $\nicefrac{v}{c}$ where $v$ is the magnitude of the velocity): in this scheme, the 2-component spinors $\phi$ and $\chi$ form the so-called \emph{small} and \emph{large components}, respectively \cite{MandlShaw}.

Substituting the small components \reff{small-components} in eq. \reff{stationary-eq1}, after standard manipulation we get 
\begin{equation}\label{generalized-pauli}
E\,\chi=
\tfrac{1}{2m}\left[p^{2}\,+\,C_0^{2}\,+\,2\,C_0\,(\sigma_i\,p^{\,i})\,+\,
\sigma_i\,C^i\right]\,\chi\;.
\end{equation}
This equation exhibits strong analogies with the electro-magnetic case. In particular, it is interesting to investigate the analogue of the so-called Pauli Equation used in the analysis of the energy levels as in the Zeeman effect \cite{bransden}: 
\begin{equation}
E\,\chi=
\left[\tfrac{1}{2m}\;(p^{2}+e^{2}\mt{A}^{2}+2e\mt{A}_ip^i)+\mu_B
(\sigma_i B^i)\,-e\,\Phi^{\scriptscriptstyle (E)}\right]\chi\;,
\end{equation}
where $\mu_B=e/2m$ is the Bohr magneton and $\mt{A}_i$ denotes the vector-potential components, $B^i$ being the components of the external magnetic filed and $\Phi^{\scriptscriptstyle (E)}$ the electric potential.

Let us now neglect the second order term $C_0^{2}$ in eq. \reff{generalized-pauli} and implement the symmetry: $\p_\mu\to\p_\mu+\mt{A}_\mu^{U(1)}+A_\mu^{\ph ab}\Sigma_{ab}$, with a vanishing electromagnetic vector potential, \emph{i.e.}, $\mt{A}_i\equiv0$. Introducing a Coulomb central potential $V(r)$ through the substitution $E\to E-V(r)$, we can derive the total Hamiltonian of the system, \emph{i.e.}, $H_{tot}=H_0+H^{\primo}$, where
\begin{align}
H_0&=\frac{p^{2}}{2m}-\frac{Ze^{2}}{(4\pi\epsilon_0)r}\;,\qquad\qq H^{\primo}=H_1+H_2\;,\\
H_1&=\;C_0\,(\sigma_i\,p^{\,i})\;/\;m\;,\quad\qq\q
H_2=\sigma_i\,C^i\;/\;2m\;,
\end{align}
which characterize the electron dynamics in an hydrogen-like atom in presence of a gauge field of the LG (here, $Z$ is the atomic number and $\epsilon_0$ denotes the vacuum dielectric constant). The solutions of the unperturbed Hamiltonian are the well-known modified two-components Schroedinger wave function, that write $H_0\,\psi_{n\,\ell\,\ml\,\ms}=E_n\;\psi_{n\,\ell\,\ml}(\textbf{r})\;\chi_{\nicefrac{1}{2},\,\ms}$, where the energy levels are $E_n=-m\,(Z\alpha)^{2}\;/\;2n^{2}$.

Since $H_1$ and $H_2$ have to be treated like perturbations, the gauge fields can be considered as independent, in the low-energy (linearized) regime. The analysis of $H_1$ can be performed substituting the operator $(\sigma_i\,p^i)$ with $(J_i p^i)$, where $J_i$ denote the components of the total angular momentum operator (in fact, $L_i p^i=0$). $H_1$ is diagonal in the basis $\basej$ and according to basic tensor analysis, we decompose the term $(J_i p^i)$ into spherical-harmonics components. As a result, by means of the Wigner-Eckart Theorem \cite{Sakurai}, we find non-vanishing matrix elements corresponding to
\begin{equation}
j^{\primo}=j+1\;,\qquad\qq\mj^{\primo}=\mj\;.
\end{equation}
Anyhow, since $(J_i p^i)$ is a pseudo-scalar operator (\emph{i.e.}, it connects states of opposite parity), no transition is eventually allowed.

The analysis of $H_2$ requires a different approach. We assume that the new Lorentz fields are directed along the $z$ direction. This way, only the component $C_3$ is considered and, for the sake of simplicity, we impose that only one between $A_1^{02}$ and $A_2^{01}$ contributes, in order to recast the correct degrees of freedom. The effect of $C_3$ corresponds to that of an external ``magnetic'' field generated by the fields $A_k^{0j}$, which can be considered the vector bosons (spin$-1$ and massless particles) of such an interaction. $H_2$ is now diagonal in the unperturbed basis $\basel$ and produce an energy-level split of the order 
\begin{equation}
\Delta E=\tfrac{C_3}{\hbar m}\; \ms\;,
\end{equation}
where $\ms=\pm\nicefrac{1}{2}$. Nevertheless, because of electric-dipole selection rules \cite{bransden}, we have to impose $\Delta\ms=0$, and no correction to the well-known transitions is detectable.

Collecting all the results together, we conclude that no new spectral line arises. Because of this properties of the Hamiltonian, it is not possible to evaluate an upper bound for the coupling constant of the interaction.

\section{Curved Space-Time and the Role of Torsion}
The considerations developed in flat space-time can be generalized to curved space-time keeping in mind that the torsion-less assumption of GR (perfectly realized by the Hilbert-Palatini Aaction) does not allow for an independent gauge field of the LG \cite{afldb}. In what follows, we provide, in the First-Order Approach, a link between the dynamics of the contortion field $\mt{K}_\mu^{\ph ab}$ and the new Lorentz gauge fields. 

The need to introduce Lorentz connections $A_\mu^{\ph ab}$ in curved space-time is motivated by the restoration of the spinor Lagrangian local invariance under diffeomorphism-induced Lorentz transformations, while spin connections $\omega_\mu^{\ph ab}$ allow one to recover the proper Dirac algebra. The generalization consists in considering a curved manifold, in which the tetrad basis is the dynamical field describing pure gravity. Local Lorentz transformations are still considered as a gauge freedom and, like in flat space, new connection fields have to be introduced.

Considering the Riemann-Cartan space $U^{4}$ filled with \emph{general} affine connections $\tilde{\Gamma}_{\mu\nu}^{\rho}$, the torsion field $\mt{T}_{\mu\nu}^{\rho}$ is defined as
$\mt{T}_{\mu\nu}^{\rho}=\tilde{\Gamma}_{[\mu\nu]}^{\rho}$. Within the framework of the First-Order Approach \cite{kim}, the II Cartan Structure Equation writes
\begin{equation}\label{Cartan eq general}
\p_\mu e^{\ph a}_{\nu} -\p_\nu e^{\ph a}_{\mu}-\tilde{\omega}_{\mu}^{\ph ab}e_{\nu b}
+\tilde{\omega}_{\nu}^{\ph ab}e_{\mu b}=
e^{\ph a}_{\rho}\,\tilde{\Gamma}_{[\mu\nu]}^{\rho}=
e^{\ph a}_{\rho}\mt{T}_{\mu\nu}^{\rho}=
\mt{T}_{\mu\nu}^{\ph\ph a}\;.
\end{equation}
The total connections $\tilde{\omega}_{\mu}^{\ph ab}$, solution of this equation, are
\begin{equation}\label{connection}
\tilde{\omega}_{\mu}^{\ph ab}=\omega_{\mu}^{\ph ab}+\mt{K}_{\mu}^{\ph ab}\;,
\end{equation} 
where $\mt{K}_{\mu}^{\ph ab}$ is the projected contortion field which is derived by the usual relation 
\begin{equation}
\mt{K}^{\mu}_{\nu\rho}=-\tfrac{1}{2}(\mathcal{T}^{\mu}_{\nu\rho}-\mathcal{T}_{\rho\nu}^{\mu}+\mathcal{T}^{\mu}_{\nu\rho})\;,
\end{equation}
while the $\omega_\mu^{\ph ab}$'s  are the standard spin connections, \emph{i.e.}, $\omega_{\mu}^{\ph ab}=e^{\ph c}_{\mu}\,\gamma^{ba}_{\ph\ph c}$.

As far as the formulation of a diffeomorphism-induced Lorentz gauge theory is concerned, new connections $A_\mu^{\ph ab}$ have to be introduced. To establish the proper geometrical interpretation of such gauge fields, let us now introduce generalized connections $\bar{\omega}_\mu^{\ph ab}$ for our model and postulate the following interaction term 
\begin{equation}\label{interacting term}
S_{conn}=2{\textstyle \int} \mathrm{det}(e)\,d^{4}x\;\;
e_{\ph a}^{\mu}e_{\ph b}^{\nu}\;\bar{\omega}_{\mu c}^{\ph [a}\,A_{\nu}^{\ph bc]}\;.
\end{equation}
In such an approach, the action describing the dynamics of the fields $A_\mu^{\ph ab}$ is derived form the gauge Lagrangian \reff{action-for-A}, \emph{i.e.},
\begin{equation}
S_A=-\tfrac{1}{4}{\textstyle \int} \mathrm{det}(e)\,d^{4}x
\;F_{\mu\nu}^{\ph\ph ab}F^{\mu\nu}_{\ph\ph ab}\;,
\end{equation}
while the action that accounts for the generalized connections can be taken as the gravitational action $S_G$ \reff{action for o}, but now the projected Riemann Tensor \reff{Riemann} is constructed by the generalized connections $\bar{\omega}_\mu^{\ph ab}$. Such a new fundamental Lorentz invariant can be denoted by $\bar{R}_{\mu\nu}^{\ph\ph ab}$ yielding
\begin{equation}\label{action-for-o1}
S_G(e,\bar{\omega})=-\tfrac{1}{4}{\textstyle \int}
\mathrm{det}(e)\,d^{4}x\;\;e^{\ph\mu}_{a}e^{\ph\nu}_{b} \bar{R}^{\ph\ph ab}_{\mu\nu}\;.
\end{equation} 

Collecting all terms together, one can get the total action for the model. Two cases can now be distinguished according to the absence or presence of spinors. If fermion matter is absent, variation of the total action wrt connections $\bar{\omega}_\mu^{\ph ab}$ gives the generalized equation
\begin{align}\label{equation for omega}
\p_\mu e^{\ph a}_{\nu} -\p_\nu e^{\ph a}_{\mu}-\bar{\omega}_{\mu}^{\ph ab}e_{\nu b}+
\bar{\omega}_{\nu}^{\ph ab}e_{\mu b}=
A_{\mu}^{\ph ab}e_{\nu b}-A_{\nu}^{\ph ab}e_{\mu b}\,,
\end{align}
which admits the solution
\begin{equation}\label{solution vacuum}
\bar{\omega}_\mu^{\ph ab}=\omega_\mu^{\ph ab}+A_{\nu}^{\ph ab}\;,
\end{equation}
As a result, confronting the expression above with the solution (\ref{connection}), the new gauge fields $A_\mu^{\ph ab}$ mimic the dynamics of the contortion field $\mt{K}_\mu^{\ph ab}$, once filed equations are considered. 

If the fermion matter contribution is taken into account in the total action, variation wrt generalized connections leads to an additional term in the rhs of eq. \reff{equation for omega}, \emph{i.e.},
\begin{equation}\label{general equation for omega}
...-\tfrac{1}{4}\,\epsilon^{ab}_{cd}\,e^{\ph c}_{\mu}e_{b\nu}\,j^{d}_{\,(ax)}+
\tfrac{1}{4}\,\epsilon^{ab}_{cd}\,e^{\ph c}_{\nu}e_{b\mu}\,j^{d}_{\,(ax)}\;,
\end{equation}
being $j_{\,(ax)}^d=\bar{\psi}\,\gamma_5\gamma^d\,\psi$ the spin axial current. The presence of spinors prevents one to identify connections $A_\mu^{\ph ab}$ as the only torsion-like components, since all the terms in the rhs of the II Cartan Structure Equation (which in this model can be identified with eqs. (\ref{equation for omega})+(\ref{general equation for omega})) have to be interpreted as torsion. This way, both the gauge fields and the spinor axial current contribute to the torsion of space-time. It is worth noting that, if the fields $A_\mu^{\ph ab}$ vanishes, we obtain the usual result of PGT \cite{blago1,blago2}, \emph{i.e.}, the Einstein-Cartan contact Theory, in which torsion is directly connected with the density of spin and does not propagate \cite{hayashi}. In our scheme, collecting eqs. (\ref{equation for omega}) and (\ref{general equation for omega}) together, we obtain the the unique solution:
\begin{equation}
\bar{\omega}_\mu^{\ph ab}=\omega_\mu^{\ph ab}+A_{\nu}^{\ph ab}
+\tfrac{1}{4}\,\epsilon^{ab}_{cd}\,e^{\ph c}_{\mu}\,j^{d}_{\,(ax)}\;.
\end{equation}
Furthermore, the spin density of the fermion matter is present in the source term of the Yang-Mills Equations for the Lorentz connections, and the Einstein Equations contain in the rhs not only the energy-momentum tensor of the matter, but also a four-fermion interaction term. The dynamical equations of spinors are formally the same as those ones of the Einstein-Cartan Model with the addition of the interaction with the LG connections $A_\mu^{\ph ab}$.

\section{Concluding Remarks}
The considerations developed in this paper have been prompted by observing that GR admits two physically different symmetries, namely the diffeomorphism invariance, defined in the real space-time, and the local Lorentz invariance, associated to the tangent fiber. Such two local symmetries reflect the different behavior of tensors and spinors: while tensors don't experience the difference between the two transformations, spinors do. In our proposal, the diffeomorphism invariance concerns the metric structure of the space-time, on the other hand, the real gauge symmetry corresponds to local rotations in the tangent fiber and admits new geometrical gauge fields. 

In our analysis, the key point has been fixing the equivalence between isometric diffeomorphisms and local Lorentz transformations. In fact, under the action of the former, spin connections behave like a tensor and are not able to ensure invariance under the corresponding induced local rotations. This picture has led us to infer the existence of (metric-independent) compensating fields of the LG which interact with spinors. Ricci spin connections could not be identified with the suitable gauge fields, for they are not primitive objects (they depend on bein vectors).

In flat space-time, we have developed the model by choosing vanishing spin connections. In treating spinor fields, a covariant derivative that accounts for the new gauge fields (behaving like natural Yang-Mills ones) has been formulated. The analysis, in flat space, is addressed considering the non-relativistic limit of the interaction between spin$-\nicefrac{1}{2}$ fields and the Lorentz gauge ones. This way, a generalization of the so-called Pauli Equation has been formulated and applied to an hydrogen-like atom in presence of a Coulomb central potential. Energy-level modifications are present but selection rules do not allow for new detectable spectral lines.

In curved space-time, a mathematical relation between the Lorentz gauge fields and the contortion field has been found from the II Cartan Structure Equation if a (unique) interaction term between the gauge fields and generalized internal connections is introduced. \vspace{0.3cm}

\textbf{*} We would like to thank Simone Mercuri for his advice about these topics.

\end{document}